\documentclass[rmp,twocolumn,showpacs,altaffilletter,superscriptaddress]{revtex4}

\usepackage{graphicx}
\usepackage{bm}
\usepackage{txfonts} 
\usepackage{relsize}
\usepackage{color}

\begin{document}

\title{Anomalous Hall effect in anatase Ti$_{1-x}$Co$_x$O$_{2-\delta}$ at low temperature regime}

\author{K.~Ueno}
\altaffiliation[Electronic mail: ]{uenok@imr.tohoku.ac.jp}
\affiliation{Institute for Materials Research, Tohoku
  University, Sendai 980-8577, Japan}

\author{T.~Fukumura}
\affiliation{Institute for Materials Research, Tohoku
  University, Sendai 980-8577, Japan}

\author{H.~Toyosaki}
\affiliation{Institute for Materials Research, Tohoku
  University, Sendai 980-8577, Japan}

\author{M.~Nakano}
\affiliation{Institute for Materials Research, Tohoku
  University, Sendai 980-8577, Japan}

\author{M.~Kawasaki}
\affiliation{Institute for Materials Research, Tohoku
  University, Sendai 980-8577, Japan}
\affiliation{
CREST, Japan Science and Technology Agency (JST), Kawaguchi 332-0012, Japan}

\newcommand{\muH}{$\mu_{\rm H}$\ }
\newcommand{\cotio }{Ti$_{1-x}$Co$_x$O$_{2-\delta}$\ }
\newcommand{\cotion }{Ti$_{0.95}$Co$_{0.05}$O$_{2-\delta}$\ }
\newcommand{\cotiox }{Ti$_{1-x}$Co$_x$O$_{2-\delta}$}
\newcommand{\cotionx }{Ti$_{0.95}$Co$_{0.05}$O$_{2-\delta}$}
\newcommand{\tio }{TiO$_2$\ }
\newcommand{\tiox}{TiO$_2$}
\newcommand{\rA }{$\rho_{\rm AHE}$\ }
\newcommand{\sA }{$\sigma_{\rm AHE}$\ }
\newcommand{\sAx }{$\sigma_{\rm AHE}$}
\begin{abstract}
Anomalous Hall effect (AHE) of a ferromagnetic semiconductor anatase \cotio thin film is studied from 10\,K to 300\,K.
Magnetic field dependence of anomalous Hall resistance is coincident with that of magnetization, while the anomalous Hall resistance
decreases at low temperature in spite of nearly temperature-independent magnetization.
Anomalous Hall conductivity \sA is found to be proportional to the square of Hall mobility,
suggesting that charge scattering strongly affects the AHE in this system.
The anatase \cotio also follows a scaling relationship to conductivity $\sigma_{xx}$\ as \sA$\propto {\sigma_{xx}}^{1.6}$, which was observed for another polymorph 
rutile \cotiox, suggesting an identical mechanism of their AHE.
\end{abstract}
\date{\today}
\maketitle
Anomalous Hall effect (AHE) commonly observed for ferromagnetic metals is a good measure of magnetic ordering for thin film specimen and
also is an important physical quantity for semiconductor spintronics devices such as field-effect transistor.\cite{Ohno_fet,Chiba_fet}
Both anatase and rutile phases \cotio are one of the most promising candidates
for future spintronics devices because of their room temperature ferromagnetism.\cite{matsumoto_science,fukumura_sst}
For rutile \cotiox, systematic variation in  AHE was observed as functions of $x$\ and
charge density, representing typical characteristics of ferromagnetic semiconductors.\cite{toyo_nmat}
Also, the magnetic tunneling junction has already been realized to show tunneling magnetoresistance up to 200\,K.\cite{toyo_jjap}
For a field-effect switching of ferromagnetism, anatase \cotio is more appropriate than rutile \cotio
because ferromagnetism appears at lower charge density for the former, making the field effect switching easier.
However, there has been only a few studies reported for AHE of anatase \cotiox.\cite{hitosugi_ahe,cho_ahe,jansen_ahe}
Therefore, detailed studies on AHE with comparing other ferromagnetic semiconductors are necessary in order to elucidate validity of field effect applications.
In addition, AHE itself is currently investigated both theoretically and experimentally since detail mechanism of AHE has not been clear so far.\cite{jung_prl,nagaosa_ahe}

\begin{figure}
\includegraphics[width=8cm]{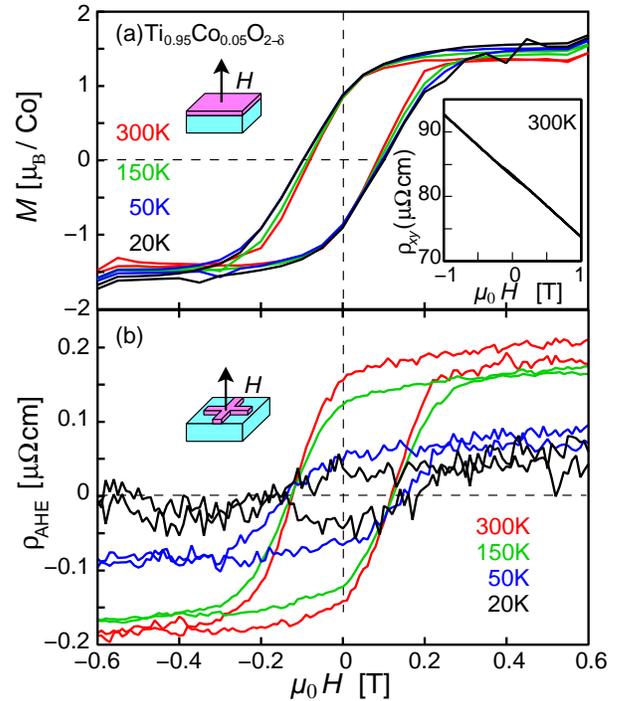}
\caption{
Magnetic field $H$\ dependence of (a) magnetization $M$\ and (b) anomalous Hall resistivity $\rho_{\rm AHE}$\ of an anatase Ti$_{0.95}$Co$_{0.05}$O$_{2-\delta}$ film at various temperatures.
Inset in (a) is $H$\ dependence of transverse Hall resistivity $\rho_{xy}$\ at 300\,K.
}
\end{figure}
Here, we report on the AHE of an anatase \cotio thin film at low temperature regime,
focusing on dependence of AHE on electric transport properties and magnetization.
Particularly, scaling relationship between anomalous Hall conductivity and conductivity for anatase \cotio is compared with anothor
polymorph rutile \cotio.

The anatase \cotion film\,(40\,nm) was grown on (001) LaAlO$_3$\ single crystal substrate by pulsed laser deposition with KrF excimer laser
in $\rm 10^{-6}$\,Torr of oxygen at 400\,$\rm{}^\circ $C after the deposition of \tio buffer layer\,(5\,nm) 
in $\rm 10^{-3}$\,Torr of oxygen at 650\,$\rm{}^\circ$C.
The reflection high-energy electron diffraction intensity was monitored 
\textit{in situ} during the film growth, and the intensity oscillation was observed
during the initial stage of growth up to 15\,nm of the \cotion layer.
The deposited film was divided into two pieces.
One was used for magnetization measurements where the magnetization data of the film was deduced from the measured magnetization data by subtracting
the diamagnetic background of the LaAlO$_3$\ substrate.
Another was photolithographically patterned into Hall bars with a channel of 200\,$\mu$m long $\times$
60\,$\mu$m wide for transport measurements.

Figure 1\,(a) shows magnetic field $H$\ dependence of the magnetization $M$\ at various temperatures.
$M$\ is almost temperature independent and saturated above 0.5\,T.
Inset of Fig.~1 shows $H$ dependence of transverse Hall resistivity $\rho_{xy}$\ at 300\,K.
In ferromagnetic metals, $\rho_{xy}$\ is the sum of ordinary and anomalous parts of Hall resistivity.
The former is expressed as $R_{\rm O}H$, where $R_{\rm O}$\ is ordinary Hall coefficient which is inversely proportional to charge density $n$\ 
and used to evaluate the Hall mobility $\mu_{\rm H}$.
The latter $\rho_{\rm AHE}$\ is supposed to be  proportional to $M$. In this study, $R_{\rm O}H$ was deduced from the slope of $\rho_{xy}$ vs.\ $H$\ curves for
$|\mu_0H|>0.5$\,T, while $\rho_{\rm AHE}$\ was deduced by subtracting $R_{\rm O}H$\ from $\rho_{xy}$.
Figure 1\,(b) shows $H$\ dependence of \rA at various temperatures.
\rA vs.\ $H$\ curves are mostly coincident with $M$\ vs.\ $H$\ curves representing that $\rho_{\rm AHE}$($H$) is proportional to $M$\,($H$).
\cite{MR_footnote}
By a close look, $H$\ dependence of \rA around the coercive field is appeared to be steeper than that of $M$.
Similar difference between \rA vs.\ $H$\ and $M$ vs.\ $H$\ curves has been often observed in other ferromagnetic semiconductors such as (Ga,Mn)As.\cite{GaAs_MH}

\begin{figure}
\includegraphics[width=8cm]{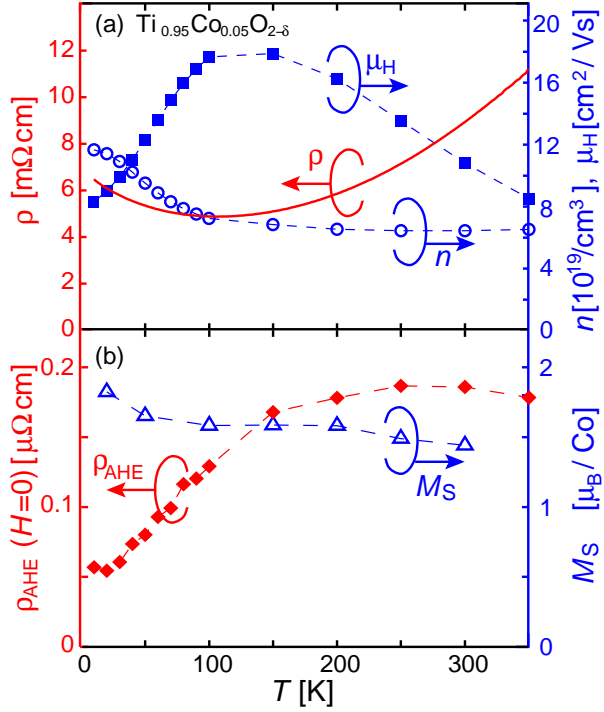}
\caption{
(a) Temperature dependence of resistivity $\rho$\,(solid line), charge density $n$\,($\circ$)
and Hall mobility $\mu_{\rm H}$\,($\blacksquare$) for the same film as in Fig.~1.
(b) Temperature dependence of anomalous Hall resistivity $\rho_{\rm AHE}$ ($\Diamondblack$) for $\mu_0H=0$\,T and
saturation magnetization $M_{\rm S}$ ($\triangle$) for $\mu_0H=1$\,T extracted from Fig.~1.
}
\end{figure}

Next, we will discuss the temperature dependence of electric properties in detail.
Figure 2\,(a) shows temperature dependence of $\rho$, $n$\ and $\mu_{\rm H}$.
The film shows $n$-type and metallic $\rho$\ down to 100\,K with a slight upturn of $\rho$\ at lower temperature,
in contrast to semiconducting behavior in rutile film  having similar value of $\rho$.
This is because anatase has higher $\mu_{\rm H}$ reaching to 10\,cm$^2$/Vs at room temperature in contrast with that of rutile\,(0.1\,cm$^2$/Vs).
$\mu_{\rm H}$ increases with decreasing temperature down to 150\,K following $T^{-3/2}$\ law, presumably due to phonon scattering mechanism.
At lower temperature, $\mu_{\rm H}$ rapidly decreases with decreasing temperature, presumably due to magnetic and/or ionized impurity scattering
by Co ions and oxygen vacancies.
In contrast, $n$\ is almost temperature independent down to 100\,K
followed by a slight increase with decreasing temperature without carrier freeze-out effect down to 2\,K. 
The temperature dependence of $n$\ indicates the charge conduction in metallic impurity band in this system
as is commonly observed in SrTiO$_3$\ single crystals.\cite{Lee_STO}
The increase of $n$\ might be originated from the increased permittivity at low temperature like rutile TiO$_2$,\cite{Rutile_e}
as is observed in reduced SrTiO$_3$.\cite{Lee_STO}

\begin{figure}
\includegraphics[width=8cm]{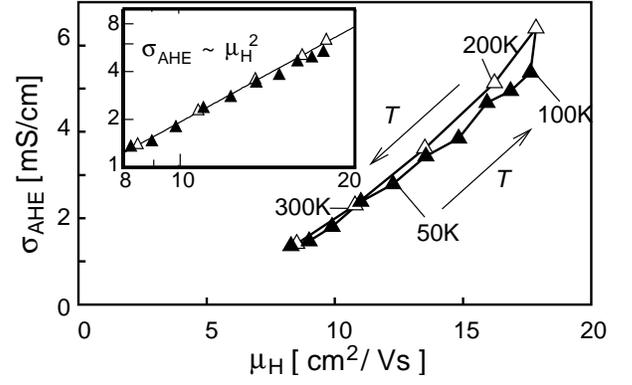}
\caption{
Hall mobility $\mu_{\rm H}$\ dependence of anomalous Hall conductivity $\sigma_{\rm \rm AHE}$.
$\sigma_{\rm AHE}$\ is deduced from $\rho_{\rm AHE}$\ and $\rho$\ values in Fig.~2
by using the relationship $\sigma_{\rm AHE} \simeq \rho_{\rm AHE} / {\rho_{xx}}^2$\,(see text).
Higher ($\ge$150\,K) and lower ($\le$ 100\,K) temperature data are represented by open and solid symbols,
respectively, for convenience.
The inset is a log-log plot of the $\sigma_{\rm AHE}$\ vs.\ $\mu_{\rm H}$\ curve.
}
\end{figure}

As shown in Fig.~2\,(b), saturation magnetization $M_{\rm S}$\ is nearly temperature independent,
whereas $\rho_{\rm AHE}$\ decreases rapidly below 150\,K with decreasing temperature,
hence $\rho_{\rm AHE}(T)$\ is not proportional to $M$\,($T$).
Such temperature variation of $\rho_{\rm AHE}$\ is related with that of transport properties in Fig.~2\,(a).
In group III-V ferromagnetic semiconductors,
Hall conductivity ($\sigma_{xy} \equiv \rho_{xy} / (\rho_{xx}^2+\rho_{xy}^2)$)\ is known to be an essential
measure of the strength of AHE,
in which ordinary Hall term is negligible.
In this study, $\sigma_{xy}$\ is approximated to be $ \rho_{xy} / \rho_{xx}^2  = R_0H/\rho_{xx}^2+\rho_{\rm AHE}/\rho_{xx}^2$\ since
$\rho_{xy}\sim\rho_{xx} / 10^3$.
Thus, we can simplify the anomalous Hall conductivity \sA as $\rho_{\rm AHE} / \rho_{xx}^2$.
Figure 3 shows the $\mu_{\rm H}$\ dependence of \sA at various temperatures obtained from the data in Fig.~2.
It appears that \sA has good one-to-one correspondence to $\mu_{\rm H}$, irrespective of temperature ranges where
$\mu_{\rm H}$\ is limited by the phonon and impurity scattering at higher and lower temperatures than 150\,K, respectively.
As shown in inset of Fig.~3, \sA is proportional to ${\mu_{\rm H}}^2$.
For single band conduction model, $\mu_{\rm H}$\ is expressed as $e\tau/m^*$ ($e$: electron charge, $m^*$: effective mass and $\tau$: scattering time) where $\tau$\ is temperature dependent parameter governed by various charge scattering mechanisms.
Taking the constant $m^*$\ into account, \sA can be connected with $\tau$\ regardless of the scattering mechanism.

\begin{figure}
\includegraphics[width=8cm]{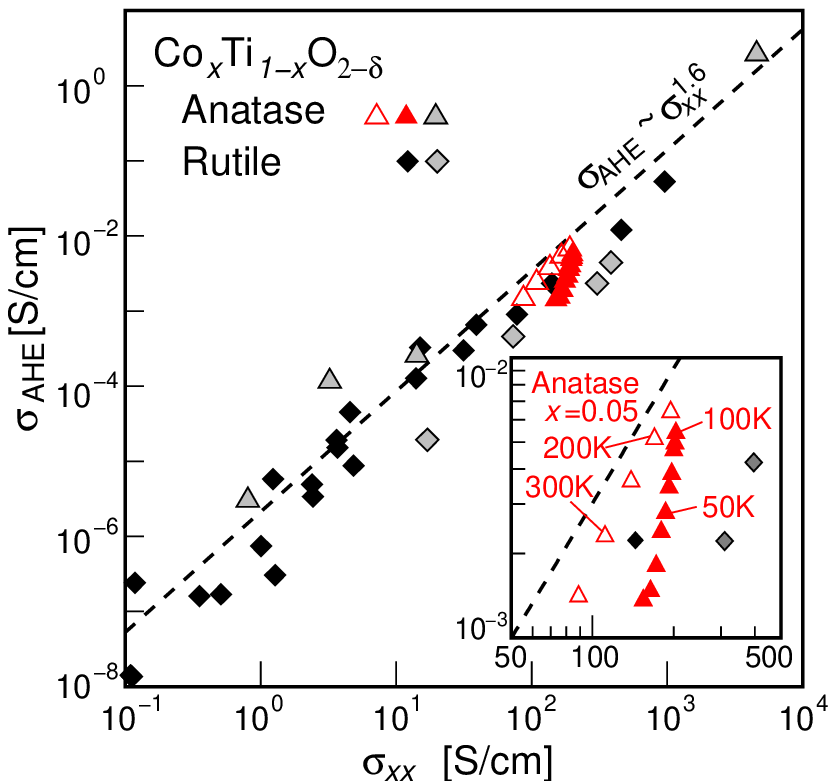}
\caption{
Relationship between the anomalous Hall conductivity $\sigma_{\rm AHE}$\ and conductivity $\sigma_{xx}$\ for
anatase \cotio (triangles.\ this work and Refs.7-9) and
rutile \cotio(diamonds.\ Refs.5 and 16).
The solid gray symbols correspond to data taken from other groups.
The inset shows a magnification around the data for anatase \cotion in this work, where the open and
solid triangles represent data for higher ($\ge$150\,K) and lower ($\le$ 100\,K) temperatures.
}
\end{figure}
Let us examine the observed AHE together with the other ferromagnetic semiconductors.
For group III-V ferromagnetic semiconductors, \sA is theoretically expressed as a function of conductivity ${\sigma_{xx}}^\alpha$,
where the exponent $\alpha$\ reflects the origin of AHE
such as skew scattering and side-jump mechanisms (corresponding to $\alpha\simeq$1 and $\simeq$0, respectively).\cite{jung_prl}
On the other hand, we reported that rutile phase \cotio has $\alpha$\ of $\sim$1.6 experimentally irrespective of temperature, $x$, and $n$.\cite{toyo_nmat}
The transport properties of anatase \cotio is qualitatively different from that of rutile \cotiox:
the former shows band conduction with large $\mu_{\rm H}$\,($\sim$10\,cm$^2$/Vs) similar to group III-V ferromagnetic semiconductors,
while the latter shows hopping conduction with small $\mu_{\rm H}$\,($\sim$0.1\,cm$^2$/Vs) similar to typical transition metal oxides.

Relationship between $\sigma_{xx}$\ and \sA for the anatase \cotion film is shown in Fig.~4 by red triangles.
Data for anatase (Ti$_{0.92}$Nb$_{0.03}$Co$_{0.05}$O$_2$ (Ref.~7), Ti$_{0.96}$Co$_x$Ni$_{0.04-x}$O$_{2-\delta}$(Ref.~8) and Ti$_{0.0986}$Co$_{0.014}$O$_{2-\delta}$(Ref.~9)) and rutile (Refs. 5, 16) phases from different research groups  are also plotted for comparison.
The data for anatase \cotio including the other reports fall into the same relationship as those for rutile \cotio with $\alpha$\ of $\sim$1.6,
irrespective of temperature, $x$, and $n$.
This relationship seems to be universal for these \cotio polymorphs in spite of the different transport properties as described above.
Recent theory indicates that the exponent $\alpha$\ depends on $\sigma_{xx}$\ and the amplitude of impurity potential.\cite{nagaosa_ahe}
The $\alpha$\ of $\sim$1.6 is reproduced in case of the weak impurity potential with $\sigma_{xx}\le10^3$\,S/cm\,(dirty limit),
while $\alpha$\ is smaller than 1 for $\sigma_{xx}$$\ge$$10^3$\,S/cm\,(clean limit).
The $\alpha\sim$1.6 relationship has been experimentally reported also for transition metal oxides (TMOs)
such as Nd$_2$(Mo$_{1-x}$Nb$_x$)$_2$O$_7$\ and La$_{1-x}$Sr$_x$CoO$_3$.
This result suggests that AHE of \cotio is within a class of such metallic TMOs in spite of the broad range of their mobility.

In conclusion, the anomalous Hall effect of anatase \cotio is shown to have almost the same magnetic field dependence as magnetization,
and to decrease with decreasing temperature in contrast with the temperature-independent magnetization.
The anomalous Hall conductivity follows the unified relationship with the mobility, $\sigma_{\rm AHE}\sim{\mu_{\rm H}}^2$,
suggesting significant role of the charge scattering on AHE.
The scaling relationship between $\sigma_{\rm AHE}$\ and $\sigma_{xx}$\ follows that of rutile \cotio reported previously in spite of the
different conduction mechanisms. The scaling exponent is explained by the recent theory and coincides with those of several transition metal oxides.

We would like to thank D.~Chiba, F.~Matsukura, N.~Nagaosa, H.~Ohno and S.~Onoda for valuable discussions
and R.~Jansen for discussions with unpublished data.
This work was supported by the MEXT Grant of Creative Scientific Research No.~14GS0204,
by the MEXT Grant of the Scientific Research on Priority Areas(16076205),
and by the NEDO, Industrial Research Grant Program (05A24020d).


\begin{thebibliography}{1}
\bibitem{MR_footnote}
Asymmetric \rA vs.\ $H$\ curve at 20\,K is accompanied by negative magnetoresistance superposed on the Hall resistance.
\end{thebibliography}
\end{document}